\documentclass[twocolumn,showpacs,preprintnumbers,amsmath,amssymb]{revtex4}

\usepackage{graphicx}
\usepackage{dcolumn}
\usepackage{bm}
\usepackage{multirow}
\usepackage{amsmath}
\def\mathbi#1{\textbf{\em #1}}

\begin{document}
\title{Influence of halide composition on the structural, electronic and optical properties of mixed CH$_3$NH$_3$Pb(I$_{1-x}$Br$_x$)$_3$ perovskites with virtual crystal approximation method}

\author{Un-Gi Jong,$^1$ Chol-Jun Yu\footnote{Corresponding author: ryongnam14@yahoo.com},$^1$ Jin-Song Ri,$^1$ Nam-Hyok Kim,$^2$ and Guk-Chol Ri$^2$}
\affiliation{$^1$Department of Computational Materials Design, Faculty of Materials Science, and $^2$Department of Theoretical Physics, Faculty of Physics, Kim Il Sung University, Ryongnam-Dong, Taesong District, Pyongyang, DPR Korea}

\date{\today}

\begin{abstract}
Extensive studies undertaken recently have demonstrated promising capability of the hybrid halide perovskite CH$_3$NH$_3$PbI$_3$ in solar cells with high power conversion efficiency exceeding 20\%. However, the existence of intrinsic and extrinsic instability in these materials remain a major challenge to commercialization. Mixing halides is expected to resolve this problem. Here we investigate the effect of chemical substitution on the structural, electronic and optical properties of mixed halide CH$_3$NH$_3$Pb(I$_{1-x}$Br$_x$)$_3$ perovskites using the virtual crystal approximation method within density functional theory. As the Br content $x$ increases from 0.0 to 1.0, the lattice constant decreases in proportion to $x$ with a function of $a(x)=6.420-0.333x$ (\AA), while the band gap and the exciton binding energy increase with a quadratic function of $E_g(x)=1.542+0.374x+0.185x^2$ (eV) and a linear function of $E_b(x)=0.045+0.057x$ (eV) respectively. The photoabsorption coefficients are also calculated, showing blue-shift of the absorption onsets for higher Br contents. We calculate the phase separation energy of these materials and analyze the electronic density difference to estimate the material stability. Based on these results, we suggest that the best match between efficiency and stability can be achieved at around $x=0.2$ in CH$_3$NH$_3$Pb(I$_{1-x}$Br$_x$)$_3$ perovskites.
\end{abstract}

\maketitle

\section{\label{sec:intro}Introduction}
Perovskites based solar cells (PSCs) are affording a promise of bright future of solar energy utilization, with the fast rise of power conversion efficiency already exceeding 20\%~\cite{Zhou14,Jeon}, remarkably easy fabrication process~\cite{Liu,Liu1,Casaluci}, and moderately low cost and sufficient supply of raw materials~\cite{Giacomo,Burschka}. The key component for governing the PSC's performance is methylammonium lead tri-iodide perovskite (CH$_3$NH$_3$PbI$_3$ or MAPbI$_3$) that is used as charge carrier mediator as well as light absorber. Numerous studies undertaken for the past five years have proven that this material has ideal properties for solar cell applications such as optimal band gap around 1.5 eV~\cite{Frost14}, large absorption coefficients~\cite{Lindblad,Even,Brivio}, weak exciton binding energy ($<$ 0.05 eV)~\cite{Hakamata,Stranks15}, high mobility of free charge carrier~\cite{Edri,Lin,Manser,Yamada}, and exceptionally large charge diffusion length over 100 $\mu$m~\cite{Stranks13}. However, it also turned out that MAPbI$_3$ suffers from the poor material stability, which represents a significant challenge on route to development of commercially viable PSCs, and yet a microscopic understanding of the degradation process remains debatable~\cite{Niu,Wang,Xiao}.

Many experiments demonstrated the significant impact of environmental factors on the degradation of the device. Niu {\it et al.}~\cite{Niu} identified the extrinsic factors causing the degradation of the perovskite film, such as moisture and oxygen, ultra violet (UV) light, and thermal effect. As suggested by Burschka {\it et al.}~\cite{Burschka}, in particular, MAPbI$_3$ can be degraded easily under humid condition, and therefore, the humidity should not be over 1\% during the device fabrication. The series of reactions for the moisture catalyzed decomposition of MAPbI$_3$ into the by-products including PbI$_2$(s), CH$_3$NH$_2$(aq), I$_2$(s), H$_2$(g) and H$_2$O, which is irreversible, were proposed~\cite{Frost14}, and corroborated by measuring the X-ray diffraction (XRD) patterns before and after exposure to moisture~\cite{Niu,Yang,Christians}, and the photothermal deflection spectroscopy~\cite{Wolf}. Under illumination of UV light as well, the deterioration of PSC was occurred owing to not MAPbI$_3$ but TiO$_2$ scaffold, which is often used as transporting layer of conduction electrons. To explain this phenomena, the hypothesis was raised such that the electrons injected into TiO$_2$ layer might be trapped in deep lying unoccupied sites~\cite{Leijtens}. Thermal effect is also a considerable factor in the stability of compound. When elevating the environmental temperature, MAPbI$_3$ may be decomposed into PbI$_2$ and CH$_3$NH$_3$I (or subsequently CH$_3$NH$_2$ and HI), as confirmed by different kinds of experiments~\cite{Dualeh,Conings}. The origin of thermal decomposition was likely to be structural defects. It was also observed that the interface with TiO$_2$ or ZnO layer plays a role in the thermal decomposition, whose mechanism was suggested to be a deprotonation of the MA cation in contact with the interface~\cite{Yang2}.

In addition to such extrinsic factors, Zhang {\it et al.}~\cite{Zhang} reported from first-principles calculations that MAPbI$_3$ is intrinsically instable. The authors calculated the energy change for the phase separation, CH$_3$NH$_3$PbI$_3$$\rightarrow$CH$_3$NH$_3$I+PbI$_2$, showing that this reaction is exothermic and thus it may occur spontaneously even without any moisture, oxygen or UV light in the environment. Once it is formed, however, the kinetic barrier may prevent the compound from phase separation, so that MAPbI$_3$ may still be stable for a certain period to be used safely in PSCs. In addition, they suggested a promising method to improve the stability of PSCs, {\it i.e.,} substitution of ingredient ions by similar elements, e.g., replacing Pb$^{2+}$ by Sn$^{2+}$, I$^-$ by Br$^-$ or Cl$^-$, or CH$_3$NH$_3^+$ by Cs$^+$. In this context, it is remarkable to tune the efficiency as well as operational stability of PSCs by adjusting the structural composition and order of mixed halide perovskite compounds~\cite{Mosconi,Noh,Aharon}.

The most favorable attempt to improve the material instability of MAPbI$_3$ is to replace I ions with Br or Cl ions~\cite{Noh,Sadhanala,Atourki}. Noh {\it et al.}~\cite{Noh} showed that increasing the Br content $x$ in MAPb(I$_{1-x}$Br$_x$)$_3$ causes a phase transition, e.g., from tetragonal to cubic at $x$=0.13, and moreover altered the band gap following the quadratic function of Br content, $E_g(x)=1.57+0.39x+0.33x^2$ (eV). For low Br concentration ($x${$<$}0.2), in particular, while the efficiency almost unchanged, the stability of device was found to be significantly improved. When exposing to a relative humidity of 55\%, the cells underwent the significant degradation for the $x$=0 and $x$=0.06 cases, but for higher Br contents ($x$=0.2 and $x$=0.29) no large degradation was observed within the 20 days measurement period. It was suggested that the improved stability for higher Br content is due to a reduced lattice constant and a phase transition from tetragonal to cubic phase. In spite of such experiments, few theoretical study based on the first-principles method to uncover the fundamental mechanism of the stability improvement by substituting Br for I can be found, to the best of our knowledge.

In this work, we investigate the mixed iodide-bromide perovskite compounds MAPb(I$_{1-x}$Br$_x$)$_3$ to address the effects of Br substitution on the material properties at the electronic scale. To conduct the first-principles simulations of solid solutions with moderate computational cost, we utilize the virtual crystal approximation (VCA)~\cite{yucj07,Iniguez} rather than the supercell method. As increasing the Br content $x$ from 0.0 to 1.0 with the interval of 0.1, the lattice parameters and band gaps are calculated and compared with the experiments to verify the validity of the underlying VCA method. Furthermore, we describe a method to approximately calculate the exciton binding energy, and with this, demonstrate the merit of charge carrier generation and transportation in these materials. Finally, we calculate the formation energy of these materials MAPbX$_3$ from their compositional phases MAX and PbX$_2$ to determine whether the reactions could be exothermic or endothermic, and the charge density difference to find out the charge transfer during the formation.

\section{\label{sec:theoretics}Method}
\subsection{\label{subsec:exciton}Exciton binding energy}
To calculate the exciton binding energy, we make use of effective mass approximation, in which an exciton made up of a hole and an electron can be viewed as a hydrogen atom. The difference between hydrogen atom and the excitonic system is attributed to the replacement of electron and nucleus masses by the effective masses of electron ($m_e^*$) in conduction band and hole ($m_h^*$) in valence band, which can be considered to be isotropic for simplicity, and to the consideration of the dielectric constant $\varepsilon$ of surrounding material in an attractive Coulomb interaction between them. Dealing with the excitonic system in the same way of solving the Schr\"{o}dinger equation of hydrogen atom gives the eigenvalues of excitonic system, from which the exciton binding energy can be provided as follows,~\cite{AMsolid}
\begin{equation}
 \label{eig_exciton}
E_b=\frac{m_ee^4}{2(4\pi\varepsilon_0)^2\hbar^2}\frac{m_r^*}{m_e}\frac{1}{\varepsilon^2}\approx13.56\frac{m_r^*}{m_e}\frac{1}{\varepsilon^2}~(\text{eV})
\end{equation}
where $m_r^*$ ($1/m_r^*=1/m_e^*+m_h^*$) is the effective reduced mass replacing the hydrogen atom reduced mass $\mu$ ($1/\mu=1/m_e+1/m_p\approx1/m_e$), and $\varepsilon_0$ is the dielectric constant of vacuum. Therefore, to calculate the exciton binding energy with this method, we need to know the dielectric constant of materials and the effective masses of electron and hole, which can be obtained readily through the first-principles calculations.

The validity of this exciton model known as {\it Mott-Wannier exciton} model can be described by estimating the extending radius of the lowest bound state like
\begin{equation}
\label{rad_exciton}
a_0^*=\varepsilon\frac{m_e}{m_r^*}a_0,
\end{equation}
where $a_0$ is the Bohr radius; $a_0^*$ must be larger than the lattice constant for this model~\cite{AMsolid}. In the case of semiconducting materials with small effective masses and large dielectric constants, it is not difficult to satisfy this requirement. When the atomic levels become more tightly bound, the exciton will become more localized (known as {\it Frenkel exciton} in the extremely localized case), due to the decrease of $\varepsilon$ and the increase of $m_r^*$.

\subsection{\label{subsec:details}Computational method}
For all relevant atoms, the optimized norm-conserving pseudopotentials with the designed nonlocal potential suggested by Rappe {\it et al}.~\cite{Rappe} were constructed using the Opium package~\footnote{The Opium package has features of atomic structure calculation, norm-conserving pseudopotential generation, and conversion into the fhi format, which is acceptable to the ABINIT package, being available at http://opium.sourceforge.net.}. The valence electronic configurations of atoms are as follows; H--1s$^1$, C--2s$^2$2p$^2$, N--2s$^2$2p$^3$, Br--4s$^2$4p$^5$, I--5s$^2$5p$^5$, and Pb--5d$^{10}$6s$^2$6p$^2$. Here, the pair of Br and I atoms, having the same valence electronic configuration, was treated as the virtual atom. To construct the pseudopotential of this virtual atom, we have utilized the {\it Yu-Emmerich extended averaging approach} (YE$^2$A$^2$ in short)~\cite{yucj07}, in which both potentials and wavefunctions are averaged. The Perdew-Burke-Ernzerhof (PBE)~\cite{pbe} formalism for exchange-correlation functional within generalized gradient approximation (GGA) was used to generate the pseudopotentials and further perform the crystalline solid simulations.

\begin{figure}[!th]
\begin{center}
\includegraphics[clip=true,scale=0.3]{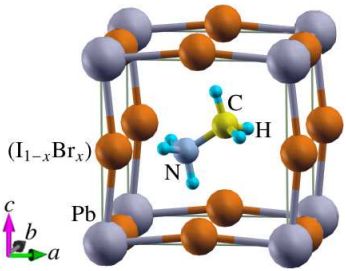}
\end{center}
\caption{\label{fig_unit}(Color online) Ball-and-stick model of unit cell of CH$_3$NH$_3$Pb(I$_{1-x}$Br$_x$)$_3$ compound with pseudo-cubic crystalline lattice and CH$_3$NH$_3^+$ cation oriented to (101) direction.}
\end{figure}

The crystalline structure of MAPb(I$_{1-x}$Br$_x$)$_3$ was assumed to be pseudo-cubic with a space group of $Pm$ as confirmed by XRD measurement~\cite{Baikie}. As shown in Figure~\ref{fig_unit}, the MA cation is oriented to the (101) direction, which is regarded as the lowest energetic configuration among the different orientations~\cite{Brivio}.

In this work, we have used the pseudopotential plane-wave method as implemented in the ABINIT (version 7.10.2) package~\cite{abinit09,abinit05}. After enough convergence tests, the plane wave cut-off energy was set to be 40 Ha and $k$-points mesh to be (4$\times$4$\times$4) for structural optimization, which guarantee the total energy convergence to be 5 meV per unit cell. For the calculations of frequency dependent dielectric constants, electronic band structure and density of states (DOS), denser $k$-point meshes (12$\times$12$\times$12) were used. To determine the optimized crystalline structure, we have calculated the total energies of the crystalline unit cells as varying the volumes evenly, at which all the atomic positions were relaxed until the atomic forces reached 0.01 eV/\AA. Then, the optimized lattice constant was determined by fitting the $E$-$V$ data into the Birch-Murnaghan equation of state.

Using the optimized unit cells, the frequency dependent dielectric constants, $\varepsilon$($\omega$)=$\varepsilon_1$($\omega$)+$i\varepsilon_2$($\omega$), were calculated within the density functional perturbation theory (DFPT)~\cite{DFPT}. Then, the photoabsorption coefficients as a function of frequency $\omega$ can be obtained using the following equation \footnote{See Ref.~\cite{Umari}, in which the authors omitted the constant $2/c$ and therefore depicted the absorbance of MAPbI$_3$ in arbitrary unit.},
\begin{equation}
\label{absorption}
\alpha(\omega)=\frac{2\omega}{c}\sqrt{\frac{1}{2}\left(-\varepsilon_1(\omega)+\sqrt{\varepsilon_1^2(\omega)+\varepsilon_2^2(\omega)}\right)}
\end{equation}
Note that the electron-hole coupling was not considered in this work, due to quite a heavy computational cost in the scheme based on the Bethe-Salpeter equation for the two-body Green's function. However, it was known that in the case of small band gap materials ignoring electron-hole coupling still leads to quite reasonable spectra compared to experiment.

\section{\label{sec:result}Results and discussion}
We first determined the lattice constants of the pseudo-cubic crystals by estimating the $E-V$ data and then fitting it into the Birch-Murnaghan equation of state. Here, the $E-V$ data was obtained by calculating the total energies of atomic-relaxed unit cells at fixed volume, as increasing the volume gradually from $0.9V_0$ to $1.1V_0$, where $V_0$ is the volume of optimized unit cell. This process was repeated at each Br content $x$, which was varied from 0 to 1 with the interval of 0.1.

\begin{figure}[!th]
\begin{center}
\includegraphics[clip=true,scale=0.5]{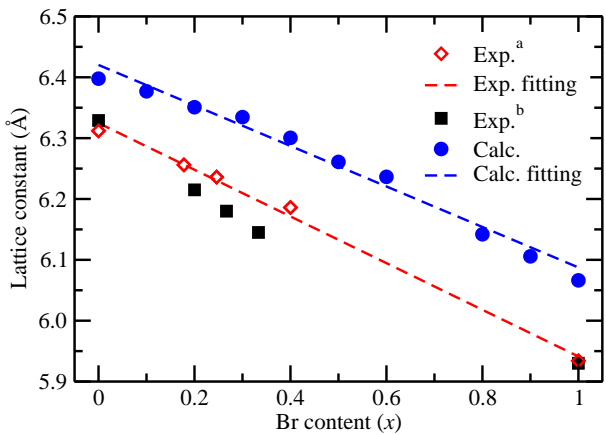}
\end{center}
\caption{\label{fig_lattice}(Color online) Lattice constants as a function of Br content $x$ in the mixed halide perovskites CH$_3$NH$_3$Pb(I$_{1-x}$Br$_x$)$_3$, with linear fitting lines (dashed lines). Red diamonds indicate the experimental values in Ref.~\cite{Misra}, filled black squares in Ref.~\cite{Atourki}, and filled blue circles mean the calculated values in this work.}
\end{figure}

In Figure~\ref{fig_lattice}, we show the optimized lattice constants as a function of Br content $x$ in MAPb(I$_{1-x}$Br$_x$)$_3$ compounds.
With the increase of the Br content, the lattice constants of the pseudo-cubic crystalline phases decrease due to the partial substitution of the larger iodine ion (ionic radius 2.2 \AA) with the smaller bromine ion (ionic radius 1.96 \AA). It is well-known that the mixed perovskites composed of two different perovskite crystals with similar lattices are expected to follow the Vegard's law, which indicates the linear dependence of lattice constants on the compositional variation. To illustrate the satisfaction of Vegard's law in the mixed halide perovskites, we have performed the interpolation of the calculated lattice constants as a linear function of Br content $x$, resulting in the formula, $a(x)=6.420-0.333x$ (\AA), which is comparable to the fitting line $a(x)=6.325-0.384x$ (\AA) to the experimental data~\cite{Misra}. Although the linear coefficient in the fitting line to the computational data is in good agreement with that to the experiment, the lattice constant at $x=0$ was overestimated the experiment with a relative error of 1.5\% and thus the line was over-shifted in $y$-axis with the same magnitude. We attribute this to the PBE-GGA exchange-correlation functional, which is in general expected to give overestimation of lattice parameters. Despite such deviation in the magnitude of lattice constants, the tangents were almost identical each other, indicating the satisfaction of Vegard's law, so that we can use our VCA method safely to draw a meaningful conclusion in the following.

\begin{figure*}[!th]
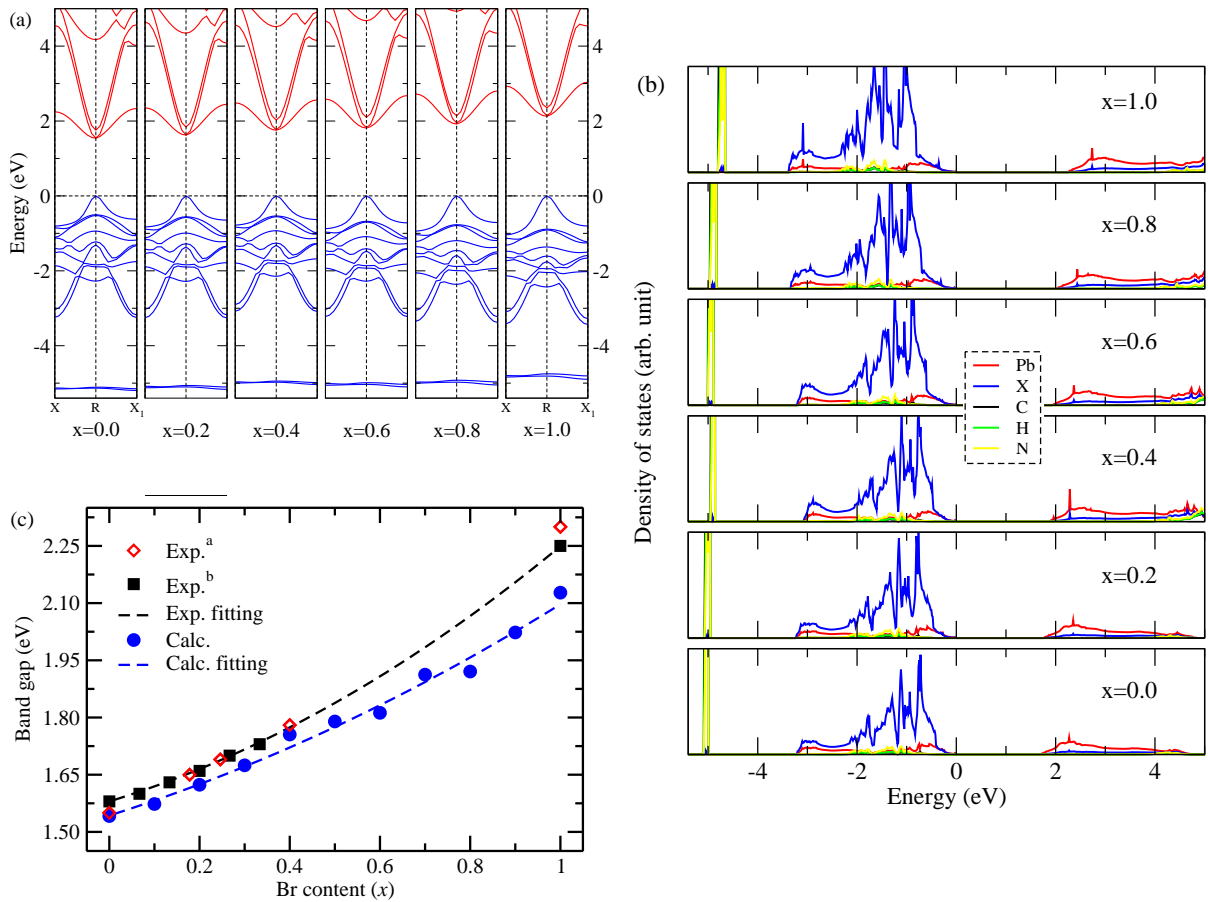

\begin{center}
\begin{tabular}{lcr}
 & & \multirow{2}{*}{\includegraphics[clip=true,scale=0.53]{fig3b.eps}} \\
\includegraphics[clip=true,scale=0.3]{fig3a.eps} & & \\
\includegraphics[clip=true,scale=0.47]{fig3c.eps} & & \\
\end{tabular}
\end{center}
\caption{\label{fig_gap}(Color online) Electronic structures of CH$_3$NH$_3$Pb(I$_{1-x}$Br$_x$)$_3$ with the variation of Br content $x$, as calculated by VCA method within DFT. (a) Electronic energy band structures around R point, (b) atomic resolved density of states, setting the top of valence band to be zero, and (c) energy band gaps as quadratic functions of Br content $x$. A virtual atom (I$_{1-x}$Br$_x$) is denoted X in (b). Exp.$^a$ and Exp.$^b$ data in (c) are from Ref.~\cite{Misra} and Ref.~\cite{Atourki}, respectively.}
\end{figure*}

We then investigated the variation tendency in electronic structures as increasing the Br content in the mixed MAPb(I$_{1-x}$Br$_x$)$_3$ perovskites, doing this with the energy band structures and partial density of states (DOS) projected on each atom. The calculated band-gaps and DOSs as functions of Br content $x$ are shown in Figure~\ref{fig_gap}. We should note that the electronic band structures and DOSs are gradually changed without any anomaly as the Br content is varied, again showing the reliability of the VCA method.

We shed light on that the band gaps are in direct mode at R point in reciprocal space over all the Br contents, shown in Figure~\ref{fig_gap} (a). This is one of the most advantageous aspects of the mixed halide perovskites MAPbX$_3$, so that the exciton (electron-hole pair) can be generated directly by light absorption without any other process like phonon. On the contrary to the case of lattice constants, the band gaps were slightly underestimated the experimental values, going further underestimation from MAPbI$_3$ ($x=0$) to MAPbBr$_3$ ($x=1$). Nevertheless, the deviations of band gaps from the experimental values, e.g., 0.01 eV at $x=0$ and 0.1 eV at $x=1$, are thought to be not very large but rather reasonably good, compared to the PBE-GGA applications to the other semiconducting compounds (the deviation $\sim$ 1 eV in general). Such a good agreement in band-gaps of these organic-inorganic halide perovskites could be explained by a fortuitous cancellation of errors, namely, the GGA underestimation is counterbalanced by the lack of spin-orbit interaction~\cite{Motta}. When occurring the substitution of heavier iodine atom ($Z=53$) by lighter bromine atom ($Z=35$), the effect of spin-orbit interaction becomes weaker, and therefore, more underestimation of band gap at MAPbBr$_3$ ($x=1$) could be expected as shown in this work.

To describe the variation tendency of band gaps with respect to the Br content, we have also interpolated the band gaps to the quadratic function of Br content $x$ in Figure~\ref{fig_gap} (c),
\begin{equation}
E_g(x)=E_g(0)+[E_g(1)-E_g(0)-b]x+bx^2,
\end{equation}
where $E_g(0)$ and $E_g(1)$ are band gaps of MAPbI$_3$ ($x$=0) and MAPbBr$_3$ ($x$=1) respectively, and $b$ is the so-called bowing parameter~\cite{Atourki}. Our calculations yielded the formula $E_g(x)=1.542+0.374x+0.185x^2$ (eV), {\it i.e.,} $E_g(0)=1.542$ eV, $E_g(1)=2.101$ eV and $b=0.185$ eV,  which are in good agreement with those from experiments $E_g(0)=1.58$ (1.579) eV, $E_g(1)=2.28$ (2.248) eV and $b=0.33$ (0.306) eV in Ref.~\cite{Noh} (Ref.~\cite{Atourki}). The bowing parameter $b$ reflects the fluctuation degree in the crystal field and the nonlinear effect arising from the anisotropic nature of binding~\cite{Atourki}. Therefore, the quite small $b$ values both in our calculation and the experiments indicate the low compositional disorder and a good miscibility between MAPbI$_3$ and MAPbBr$_3$. It is clear that the substitution of larger iodine ion by smaller bromine ion leads to the enhancement of interaction between Pb and X atoms in corner-sharing PbX$_6$ octahedra, which play a major role in determining the band structure~\cite{Even,Green}, and thus the decrease of lattice constant, resulting in the increase of band gaps with the implication of worsening the light harvesting properties from MAPbI$_3$ to MAPbBr$_3$.

We have done the analysis of the atomic resolved (and partial) DOSs in detail to seek the electronic factors possibly responsible for the band gap variation. As shown in Figure~\ref{fig_gap} (b), it is established that the valence band maximum (VBM) of MAPbX$_3$ arises predominantly from the $p$ state of X with a small contribution from the $6s$ state of Pb, while the conduction band minimum (CBM) comes from Pb $6p$ state and X $d$ and $p$ states~\cite{Hakamata}. It can be thought that Br $4p$ states tend to hybridize more strongly with Pb $s$ state than I $5p$ states, causing the down-shift of VBM and thus the increase of band gap. At this point, it is worth noting that, although the highest occupied molecular orbitals (HOMO) of MA cation are found deep $\sim$ 5 eV below the VBM, having narrow features, there is an interaction between organic MA cation and the inorganic PbX$_6$ octahedra by possible hydrogen bonding between ammonium group and X atom. It is also interesting to notice that, when increasing the Br content, the energy interval between the HOMO of MA cation and the bottom level of PbX$_6$ is getting close, providing an indication of becoming stronger MA-PbX$_6$ interaction.

\begin{figure}[!th]
\begin{center}
\includegraphics[clip=true,scale=0.5]{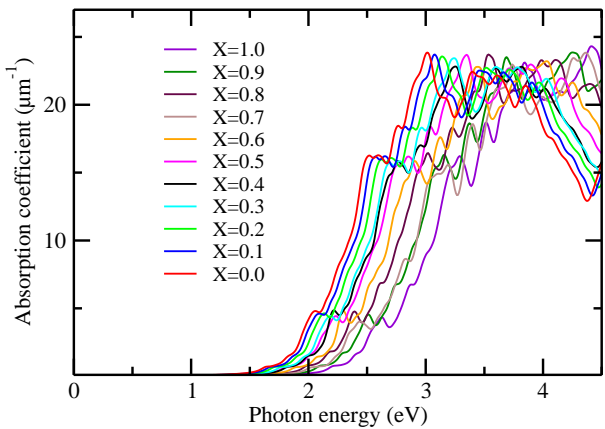}
\end{center}
\caption{\label{fig_absorption}(Color online) Photoabsorption coefficients of mixed halide perovskites CH$_3$NH$_3$Pb(I$_{1-x}$Br$_x$)$_3$ with different Br content $x$, as calculated by VCA and DFPT method within DFT.}
\end{figure}

In order to directly estimate the light harvesting capability of PSCs, we next describe the photoabsorption coefficients of these mixed halide perovskites with different Br contents, which can be obtained from Equation (\ref{absorption}) using the real and imaginary parts of frequency dependent dielectric constants, calculated within DFPT~\cite{DFPT}. In Figure~\ref{fig_absorption}, we present the photoabsorption coefficients to be functions of photon energy as increasing the Br content. For lower Br content, the mixed halide perovskites MAPb(I$_{1-x}$Br$_x$)$_3$ exhibit the extended absorption character over the whole visible light spectrum, which is an advantageous property for light harvesting. For higher Br contents, however, the absorption onset gradually shifts to the higher photon energy, {i.e.,} to shorter wavelength light. Such blue-shift of the absorption onsets can be readily expected from the rise of band gaps in the mixed halide perovskites MAPb(I$_{1-x}$Br$_x$)$_3$ with the increase of Br content.

Another important properties to be unavoidably considered in the mixed halide perovskites are the binding energy between electron and hole, which are generated due to the photon absorption, and the mobility of these charge carriers. Of two properties, the exciton binding energy plays a key role in discriminating whether the charge carriers behave free particles as in normal inorganic thin-film semiconductors, or bound excitons as in organic semiconductors. The weaker exciton binding energy indicates more freely behaving charge carriers. After the computation of the effective masses of electrons and holes by straightforward numerical process of the refined energy band data around R point in the first Brillouin zone, and the extraction of the dielectric constants simply from the frequency dependent dielectric constants, we used Equation (\ref{eig_exciton}) to calculate the exciton binding energy.

\begin{figure}[!th]
\begin{center}
\includegraphics[clip=true,scale=0.46]{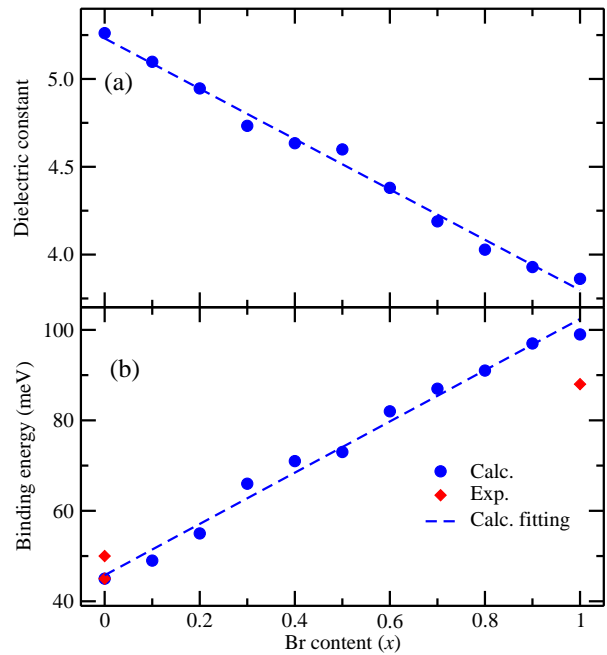}
\end{center}
\caption{\label{fig_bind}(Color online) (a) Dielectric constants and (b) exciton binding energies in mixed halide perovskites CH$_3$NH$_3$Pb(I$_{1-x}$Br$_x$)$_3$ with different Br content $x$. Experimental data is from Refs.~\cite{Sum,Innocenzo,Moses}.}
\end{figure}

In Figure~\ref{fig_bind}, we show the dielectric constants and exciton binding energies with different Br content $x$. First, let us see the effective masses of electron and hole, indirect estimation for the mobility of carriers. At $x=0$ ($x=1$), $m_e^*/m_e=0.18$ (0.21), $m_h^*/m_e=0.19$ (0.23), and $m_r^*/m_e=0.09$ (0.11). For MAPbI$_3$, these values are comparable to those of 0.12, 0.15 and 0.09 from experiment~\cite{Tanaka}, and 0.17, 0.25, and 0.11 from spin-orbit coupling GW calculations~\cite{Umari}. It is worthy to note that the effective reduced mass ($m_r^*$) slightly increases from $x=0$ to $x=1$, although there is some numerical fluctuation in the middle. As shown in Figure~\ref{fig_bind} (a), the dielectric constants decrease monotonically as increasing $x$ with a linear function of $\varepsilon(x)=5.23-1.43x$. As mentioned above, these are extracted at high frequency from the optical frequency dielectric constants, which were obtained by DFPT calculation. For MAPbI$_3$, the calculated value of 5.3 is in reasonable agreement with the recent theoretical value of $\sim$4 calculated by quantum molecular dynamics method~\cite{Hakamata}, 5.6$\sim$6.2 by PBEsol DFT calculation~\cite{Brivio}, and the experimental values of 7$\sim$10 at 10$^{12}$ Hz~\cite{Stranks15}. Dealing with these values gives the exciton binding energy of 45 meV, with an effective Bohr radius ($a_0^*$) 3 nm, which is in good agreement with the experimental value of $\sim$50 meV~\cite{Innocenzo,Miyata} and 19$\sim$45 meV~\cite{Sum}. For MAPbBr$_3$, it was calculated to be 99 meV with 1.9 nm, comparable to experimental value of $\sim$88 meV~\cite{Moses}. As discussed in Refs.~\cite{Grancini,Xiao} and looking at the not large effective Bohr radii, however, this model is valid only if the Coulomb interaction between the electron and hole is strong enough so that the lattice dielectric response does not screen the interaction. If the exciton is weakly bound, the low-frequency component of the dielectric constant (static) arising from the lattice contributions must be also considered. Brivio {\it et al.}~\cite{Brivio} reported the static dielectric constants as up to 37. This results in the decrease of binding energy down to 1 meV and the increase of scope up to 20 nm. Since the aim of this work is to investigate the variation tendency of properties as the change of Br content, we will rely on the values close to the experiment.
Based on the data shown in Figure~\ref{fig_bind} (b), we can say that the exciton binding energy increases in proportion to the Br content $x$ in MAPb(I$_{1-x}$Br$_x$)$_3$, with the linear function of $E_b(x)=0.045+0.057x$ (eV). For low Br content, in particular, the exciton binding energies are quite small, being comparable to those of the inorganic thin-film semiconductors ($<50$ meV), and therefore, the charge carriers are likely to behave free-like. Meanwhile, those are large for higher Br contents, indicating that, when increasing the Br content, the excitons become to be strongly bound.

To sum up the points so far, we can conclude that the substitution of iodine atom by bromine atom in mixed halide lead perovskites causes a little loss of advantageous properties of MAPbI$_3$ towards PSC application in overall. Then, what about the material stability?

\begin{figure}[!th]
\begin{center}
\includegraphics[clip=true,scale=0.065]{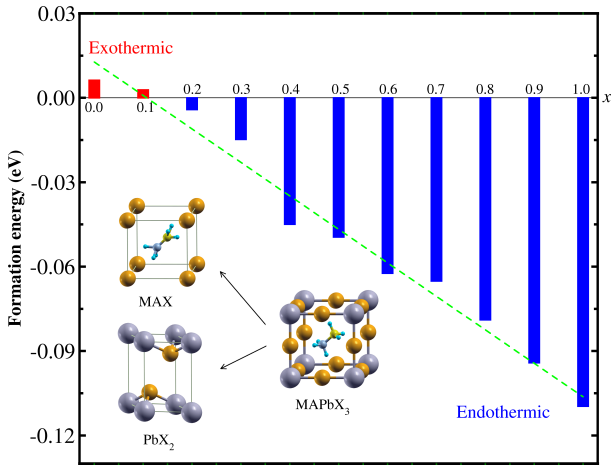}
\end{center}
\caption{\label{fig_energy_costs}(Color online) Formation energy of the mixed halide perovskites CH$_3$NH$_3$PbX$_3$ (X=I$_{1-x}$Br$_x$) from their components CH$_3$NH$_3$X and PbX$_2$ as increasing the Br content $x$. Green-colored dotted line shows linear fitting.}
\end{figure}

To make an answer to this question, we have considered the phase separation reactions in these mixed halide perovskites, MAPbX$_3$$\rightarrow$MAX+PbX$_2$. The energies required for the phase separation reactions or formation energies of MAPbX$_3$ from their compositions MAX and PbX$_2$ were calculated as the total energy difference per formula unit (f.u.), $E_f=E_\text{tot}$(MAPbX$_3)-(E_\text{tot}$(MAX)$+E_\text{tot}$(PbX$_2))$, according to the change of Br content. The crystalline structures of MAX and PbX$_2$ were assumed to be rocksalt phase with space group $Fm$\={3}$m$ and trigonal phase with space group $P$\={3}$m1$~\cite{Zhang}, respectively. Since the formation energy is in general obtained by calculating the difference between the total energies, the accuracy could be enhanced; 0.1 meV/f.u. in this work. As shown in Figure~\ref{fig_energy_costs}, the calculated formation energies at $x=0.0$ and $x=0.1$ are positive, while over $x=0.2$ they are all negative with a feature of gradual increase of magnitude when increasing the Br content $x$.
This result directly indicates that in the former cases the phase separation reactions are exothermic, while in the latter cases they are endothermic. Therefore, the phase separation of MAPbI$_3$ into MAI and PbI$_2$ may occur spontaneously without application of any extrinsic factors such as moisture, oxygen, exposure to UV light and elevated temperature, but the phase separation of MAPbBr$_3$ is thermodynamically unfavorable. This reasonably coincides with other PBE-GGA calculation~\cite{Zhang}. Moreover, Andrei {\it et al.}~\cite{Andrei} reported that time evolution of the main XRD peak of MAPbI$_3$ shows the PbI$_2$ peak as soon as MAPbI$_3$ is formed, while MAPbBr$_3$ does not give any evidence of PbBr$_2$ peak. From these results, it is unraveled that the intrinsic stability of MAPb(I$_{1-x}$Br$_x$)$_3$ is enhanced systematically as increasing the Br content, and most interestingly, the phase separation reaction becomes endothermic over $x\approx0.2$.

\begin{figure}[!th]
\begin{center}
\includegraphics[clip=true,scale=0.2]{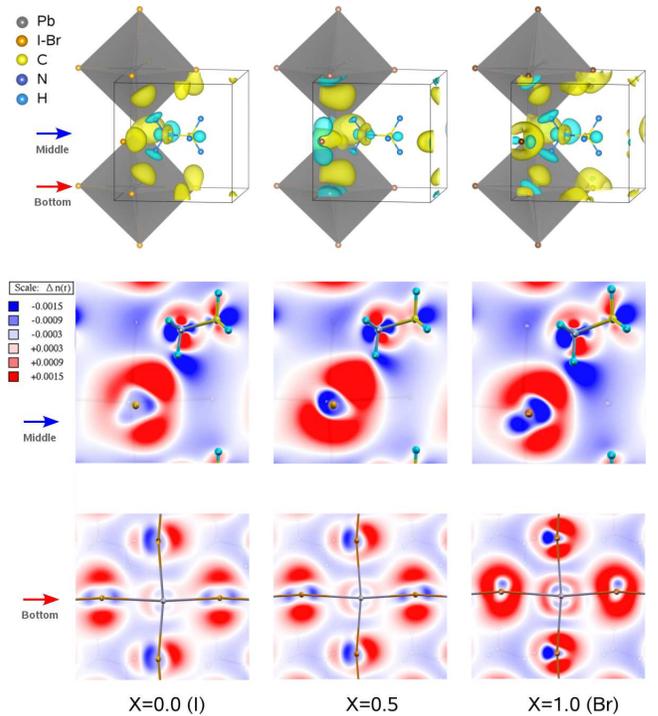}
\end{center}
\caption{\label{fig_dendiff}(Color online) Electronic density difference of the mixed halide perovskites CH$_3$NH$_3$Pb(I$_{1-x}$Br$_x$)$_3$ at the Br contents $x=0.0, 0.5, 1.0$. Upper panel shows the 3D isosurfaces evaluated at the value of $\pm$0.0015 $|e|$/\AA$^3$, where yellow (blue) color represents the positive (negative) value indicating electron accumulation (depletion), and middle and lower panels show the 2D isoline pictures on the middle plane containing CH$_3$NH$_3^+$ cation and halogen atom and on the bottom plane containing Pb and halogen atoms.}
\end{figure}

In order to get intuitive insights into the charge transfer in the event of compound formation and thus chemical bonding characteristics, the electronic density differences were calculated as the difference between the electron density of MAPbX$_3$ and those of PbX$_3$ framework and MA molecule like $\Delta n(\mathbi{r})$=$n_{\text{MAPbX}_3}(\mathbi{r})$-$(n_{\text{PbX}_3}(\mathbi{r})$+$n_{\text{MA}}(\mathbi{r}))$. In Figure~\ref{fig_dendiff}, we show the electronic density differences at $x=0.0, 0.5, 1.0$ for clear comparison. In the middle panel of Figure~\ref{fig_dendiff}, it is observed that there is much charge accumulation in the interspatial region surrounding halogen atom and charge depletion region around the hydrogen atom of MA cation and in the vicinity close to halogen atom. In the lower panel, meanwhile, great and weak charge accumulation are also seen around halogen atom and Pb atom, and charge depletion is occurred in the localized region close to halogen atom. These indicates that electrons are in general transferred from MA cation and Pb atom to halogen atom, making the chemical bond strong. We can see also that the charge transferring is gradually enhanced going from I to Br in overall, indicating that the chemical bond becomes stronger from I to Br. This reminds that the lattice constants decrease with the increase of Br content.

These computational results are correlated with the experimental findings that the compounds at $x=0.2$ or $x=0.29$ remain quite stable for a long date, whilst at $x=0.0$ and $x=0.06$ severe degraded within a short date~\cite{Noh,Wang}. Since the efficiency is expected not to be much spoiled at $x=0.2$ with the band gap of 1.62 eV and the exciton binding energy of 55 meV in this work and as confirmed with the experiment~\cite{Noh}, it can be suggested that the best match between efficiency and stability can be realized at $x\approx0.2$ in the mixed halide perovskites MAPb(I$_{1-x}$Br$_x$)$_3$.

\section{\label{sec:con}Conclusion}
Despite the remarkable advances in the performance of hybrid halide perovskites, yet the degradation and instability in these materials remain barrier to the practical use in solar cell applications. In this work, using the VCA method within DFT, we have investigated the influence of halide composition on the structural, electronic, and optical properties of the mixed halide perovskites MAPb(I$_{1-x}$Br$_x$)$_3$. When increasing the Br content $x$ from 0.0 to 1.0, we have found the decrease of lattice constants with the linear function of $a(x)=6.420-0.333x$ (\AA), while the increase of band gaps and exciton binding energies with the quadratic function of $E_g(x)=1.542+0.374x+0.185x^2$ (eV) and the linear function of $E_b(x)=0.045+0.057x$ (eV) respectively. The increase of band gaps with the Br content is due to the stronger hybridization of Br $4p$ states with Pb $s$ states than I $5p$ states, which leads to the down-shift of VBM, together with the decrease of lattice constant. With the increase of the Br content, the energy interval between the HOMO of MA cation and the bottom level of PbX$_6$ is getting close, providing an indication of becoming stronger MA-PbX$_6$ interaction. The calculated photoabsorption coefficients exhibit the blue-shift of absorption onsets for higher Br content. The substitution of I atom by Br atom leads to the enhancement of stability, which is described by analysing the energy change of phase separation reaction and electronic density difference. In conclusion, our work suggests that, considering the tunability of material properties by adjusting the Br content $x$ in the mixed halide perovskites MAPb(I$_{1-x}$Br$_x$)$_3$, the best match between efficiency and stability might be achieved at $x\approx0.2$.

\section*{\label{ack}Acknowledgments}
This work was supported partially from the Committee of Education, Democratic People's Republic of Korea, under the project entitled ``Strong correlation phenomena at superhard, superconducting and nano materials'' (grant number 02-2014). The simulations have been carried out on the HP Blade System c7000 (HP BL460c) that is owned and managed by the Faculty of Materials Science, Kim Il Sung University.

\bibliographystyle{apsrev}
\bibliography{Reference}

\end{document}